# Power law scaling of lateral deformations with universal Poisson's index for randomly folded thin sheets

Alexander S. Balankin, Didier Samayoa Ochoa, Ernesto Pineda León,

Rolando Cortes Montes de Oca, Antonio Horta Rangel, and Miguel Ángel Martínez Cruz

*Grupo "Mecánica Fractal", Instituto Politécnico Nacional, México D.F., México 07738*

We study the lateral deformations of randomly folded elasto-plastic and predominantly plastic thin sheets under the uniaxial and radial compressions. We found that the lateral deformations of cylinders folded from elasto-plastic sheets of paper obey a power law behavior with the universal Poisson's index $\nu = 0.17 \pm 0.01$, which does not depend neither the paper kind and sheet sizes (thickness, edge length), nor the folding confinement ratio. In contrast to this, the lateral deformations of randomly folded predominantly plastic aluminum foils display the linear dependence on the axial compression with the universal Poisson's ratio $\nu_e = 0.33 \pm 0.01$. This difference is consistent with the difference in fractal topology of randomly folded elasto-plastic and predominantly plastic sheets, which is found to belong to different universality classes. The general form of constitutive stress-deformation relations for randomly folded elasto-plastic sheets is suggested.

# I. INTRODUCTION

Randomly crumpled thin matter, such as polymerized membranes, biological cells, and thin sheets are of noteworthy importance for many branches of science and technology [1]. These materials exhibit three distinct phases: the flat, the tubular, and the crumpled (folded) [2]. In the last decade randomly folded materials become a subject of great interest because of their fascinating topological and mechanical properties [3,4,5,6,7,8]. The last are governed by the topology of crumpled configuration [8,9]. Specifically, the diameter of randomly folded membrane (sheet) is found to scales with the hydrostatic folding force $F$ as $R \propto F^{-\delta}$ [8], where the folding force exponent $\delta$ is universal and equal to $\delta = 3/8$ for phantom and $\delta = 1/4$ for self-avoiding membranes with a finite bending rigidity [8]. Further, it was found that the statistical topology of randomly folded elasto-plastic and predominantly plastic materials are characterized by different sets of universal scaling exponents [6].

One of the most intriguing mechanical phenomena is that an almost all deformed materials exhibit dimensional changes in lateral directions without corresponding stresses [10]. This phenomenon, known as the Poisson's effect [10, 11], may be treated as a particular expression of the Le Chatelier's principle, which states that when a system at equilibrium is subject to a change, the system will respond to relieve the effect of that change [12]. Indeed, under the equilibrium conditions, the lateral deformations lead to decrease the volume change produced by the applied strains. Specifically, in an axially loaded specimen, the Poisson's effect is commonly characterized by the ratio of lateral strain $\varepsilon_\perp$ to axial strain $\varepsilon_\parallel$, known as the Poisson's ratio,

$$\nu_e = -\varepsilon_\perp / \varepsilon_\parallel, \tag{1}$$

which is one of the fundamental physical properties of any natural or engineering material [10]. In the case of radial compression, the theory of isotropic elasticity predict that

$$\varepsilon_H = -\frac{2\nu_e}{1-\nu_e}\varepsilon_R, \tag{2}$$

where $\varepsilon_R$ and $\varepsilon_H$ are the radial and axial strains, respectively [13].

Although the Le Chatelier's principle implies a positive values of $\nu_e$, such that the Poisson's strains lead to decrease in the relative volume change, a negative Poisson's ratio (that is, a lateral extension in response to stretching) is not forbidden by thermodynamics [11]. In the limit of infinitesimally small strains, the theory of isotropic elasticity allows Poisson's ratios in the range from -1 to 0.5 for three-dimensional materials and from -1 to 1 for two-dimensional structures [13]. Materials with $\nu_e = 0$ do not exhibit changes in lateral directions. For most isotropic materials the Poisson's ratio is positive, being close to 0.15 for most ceramics, around of 0.3 for most metals and about of 0.5 for rubbery materials [10]. Generally, the values of Poisson's index and ratio are dependent on the material structure, chemical composition and porosity [14,15]. However, some classes of materials are characterized by the Poisson's ratio or index determined by the material topology. Specifically, the Poisson's ratio (index) of elastic fractals is determined by their fractal dimension [16,17,18]. Materials which expand transversely when stretched longitudinally are called auxetic [19]. The auxetic effect is usually brought about by an in-folding (reentrant) or rotating structure at either the macro- or microscopic level [20]. Examples of auxetic materials in which a negative Poisson ratio is accounted to their (micro)structure include reentrant foams, crumbled polymerized membranes, fiber networks near the percolation threshold among others (see [21,22,23,24,25,26]).

In this way, the auxetic nature of the Poisson's effect in the stretching of randomly crumbled paper was demonstrated in [27]. More generally, it was shown that the flat phase of fixed-connectivity membranes provide a wide class of auxetic materials with the universal negative Poisson's ratio $\nu \cong -1/3$, where the symbol $\cong$ denotes the numerical equality [1,26]. However, the Poisson's effect in the crumpled (folded) phase still remains ununderstood. Accordingly, in this work we studied the Poisson's effect in randomly folded elasto-plastic and predominantly plastic thin sheets subjected to the axial and radial compressions [28].

## II. EXPERIMENTS AND DISCUSSION

To study the Poisson's effect in randomly folded matter in this work we used five papers of different thickness ($h = 0.024$, $0.030$, $0.039$, $0.068$, and $0.087$ mm) and bending rigidity, early used in [6] to study the statistical topology of folded configurations. Square paper sheets with the edge size $L$ of 250, and 500 mm were folded in hands into approximately spherical balls. Then these balls we confined in cylindrical sells of different diameters to approximately cylindrical form with the height approximately equal to diameter. It should be pointed out that once the folding force is withdrawn, the sizes of folded sheet increase logarithmically with time during approximately one week, due to the strain relaxation (see, for details, refs. [6]). Accordingly, to obtain the cylinders of different dimensions from the sheets of the same size, the folded cylinders were kept in cells under hydrostatic compressions during different times (from 5 min to 48 hours) and than relaxed (during 7 days) up to no changes in the ball dimensions were observed. Then, the averaged diameter $R_0$ and height $H_0$ of each cylinder were determined from 15 random measurements. Thus we obtained the cylinders with $H_0 \cong R_0$ folded from sheets of the same size $L$ with different contraction ratios $K = L/R_0$, such that $\max K(L)/\min K(L) \geq 2$. In total, 350 folded cylinders were tested under axial compression and 50 under radial compression

Further, we performed some experiments with randomly folded aluminum foils of thickness 0.02 mm with edge sizes of 200 and 500 mm, the deformations of which are predominantly plastic [6]. Ten sheets of each size were folded in hands into approximately spherical balls with averaged diameter $R_0 = L/K$ for different contraction ratios.

**2.1. Experiment details**

First of all, folded sheets were tested under uniaxial compression in a universal testing machine (see Fig. 1 a, b). The perimeter $P(\lambda_\parallel) = 2\pi R$ of deformed cylinder (ball) was measured at different compression ratios $\lambda_\parallel = H/H_0$ with the help of silk strings. It should be pointed out that the Poisson's ratio is strictly defined only for a small strain linear elastic behavior and it is generally highly strain dependent at larger deformations [29], and even may change the sign [30]. Accordingly, in the case of large deformations, the relative lateral expansion/contraction ($\lambda_\perp = R/R_0 = P/P_0$) is commonly described by the Poisson's function of axial compression/stretching ($\lambda_\parallel = H/H_0$), where $H_0$, $R_0$, $P_0$ and $H$, $R$, $P$ are the specimen height, diameter, and perimeter before and after the deformation, respectively. The form of Poisson's function $\lambda_\perp = f_\nu(\lambda_\parallel)$ is material dependent [31] and generally, it can not be characterized by a single parameter, such as Poisson's ratio. However, for some materials the Poisson's function obeys a power law scaling behavior [16,32]:

$$\lambda_\perp = \lambda_\parallel^{-\nu} \tag{3}$$

with the strain independent scaling exponent $\nu$, called the Poisson's index, which coincides with the Poisson's ratio $\nu_e$ only in the limit of infinitesimally small strains $\varepsilon_{ii} = \sqrt{|\lambda_i^2 - 1|} \ll 1$.

The scaling behavior (3) implies that the mass density $\rho \propto V^{-1}$ of axially deformed material exhibits a fractal-like behavior, $\rho \propto \lambda_\parallel^{2\nu-1}$. So, the scaling behavior (3) is expected for soft materials with statistically self-similar structures [16,17,32]. Early, Gomes et al. [32] have reported that the lateral deformations of randomly folded aluminum foils obey the scaling relation (3) with different $\nu$ for balls folded from foils of different thickness. So, in this work we used the scaling relation (3) to determine the Poisson's index.

It should be pointed out that the deformations of folded sheets are essentially irreversible due to the plastic deformations of sheet in the crumpling creases (see Fig. 2 a, b and refs. [4-7]). Moreover, at a fixed compression ratio $\lambda_\parallel$, the compression stress slowly decreases in time during more than 24 hours (see Fig 2 c [33]). Generally, the lateral deformations of visco-elastic materials are also expected to change in time, and so, the Poisson's index (ratio) is expected to be time-dependent (see [34]]). Accordingly, to detect a possible time dependence of the Poison's expansion, in the first five experiments performed in this work the perimeter of deformed cylinder was measured several times during 24 hours for each fixed $\lambda_\parallel$. However, no time dependence in the lateral expansion was noted (see Fig. 2 d [35]).

Additionally, 50 randomly folded paper sheets [36] were tested under radial compression in the piston ring compressor (see Fig. 1 b). In this case, the axial extension ratio $\lambda_H = H/H_0$ was measured as a function of radial compression ratio $\lambda_R = R/R_0$. Notice, that for large deformations the form of dependence $\lambda_H = F(\lambda_R)$ depends on the form of constitutive equations.

**2.2. Poisson's expansion of randomly folded sheets under uniaxial compression**

In Fig 3 a the averaged perimeters of axialy deformed cylinders folded with different contraction ratios $K = L/R_0$ ($H_0 \cong R_0$) are plotted versus their heights $H = \lambda_\| H_0$. One can see that, at least in the range of $0.2 \leq \lambda_\| \leq 1$, the perimeter scales with $H$ as $P \propto H^{-\nu}$ with the same scaling exponent

$$\nu = 0.17 \pm 0.01 \qquad (4)$$

for all folded sheets tested. Accordingly, Fig. 3 b shows the data collapse in coordinates $\lambda_\perp$ versus $\lambda_\|$ for 350 folded papers sheets tested in this work under uniaxial compression. These data suggest that the Poisson's index of randomly folded paper does not depend nether on the environmental conditions (temperature and air humidity [37]), which were varied in a wide range during the six months of experiments, nor on the paper thickness, sheet size, and contraction ratio. This finding, together with the universality of local fractal dimension $D_l = 2.64 \pm 0.05$ [6], suggests the existence of universality class of crumpled phase of randomly folded elasto-plastic sheets.

At the same time, we noted that the universal value (4) differs from values of Poisson's index $\nu = 0.26 \pm 0.02$ and $\nu = 0.27 \pm 0.04$, reported in [32] for randomly folded aluminum foils of thickness 0.007 mm and 0.037 mm, respectively. So, in this work, we performed twenty experiments with randomly folded aluminum foils of thickness 0.02 mm with edge sizes of 200 and 500 mm. Surprisingly, we noted that while the lateral deformations of randomly folded foils can be fitted by the power-law relation (4) with $\nu = 0.25 \pm 0.04$ (see Fig. 4 a), close to the values reported in [32], the experimental data can be better fitted by the linear relationship (1) with the constant Poisson's ratio,

$$\nu_e = 0.33 \pm 0.01, \tag{5}$$

which is found to be independent on the foil size and the folding contraction ratio within a wide range of $\varepsilon_\parallel$ (see Fig. 4 a, b). The linear relation between lateral and longitudinal strains is quite surprising taking into account that the stress-strain relation is strongly nonlinear for $\varepsilon_\parallel \geq 0.5$ (see Fig. 2 b).

This difference in the Poisson's effect, together with the difference in the fractal dimension of randomly folded elasto-plastic and predominantly plastic sheets (see [6]), suggests that the randomly folded plastic foils and elasto-plastic sheets belong to different universality classes.

## 2.3. Radial compression test and general form for constitutive stress-deformation relationship of randomly folded sheets in one- and two-dimensional stress states

In tests with radial compression we found that the axial extension of randomly folded paper scales with the radial compression as

$$\lambda_H = \lambda_R^{-2\alpha}, \qquad (6)$$

where the scaling exponent

$$\alpha = 0.20 \pm 0.03 \qquad (7)$$

is found to be the same for all tested sheets (see Fig 3 c). So, at least numerically,

$$\alpha \cong \frac{\nu}{1-\nu}. \qquad (8)$$

where $\nu = 0.17 \pm 0.02$ is the universal Poisson's index (4) found in axial compression tests.

The scaling relations (3), (6) together with the equality (8) imply that the constitutive stress-deformation relationship of randomly folded sheets should have the following general form

$$\lambda_i^{-1} = f(\sigma_i) + \left(\lambda_j \lambda_k\right)^\alpha, \tag{8}$$

where $f(\sigma_i)$ is an increasing function of the principal stress $\sigma_i$, such that $f(0) = 0$ and $\alpha$ is defined by the relationship (5); indexes $i \neq j \neq k$ take the values 1, 2, 3, corresponding to the direction of principal stress (see Ref. [13])..

## III. CONCLUSIONS

We found that the Poisson's expansion of randomly folded elasto-plastic sheets under axial and radial compression displays a power-law scaling behavior with the universal scaling exponent (4), whereas the lateral strains of randomly folded aluminum foils depend linearly on the longitudinal strains within a surprisingly wide range of strains. The Poisson's ratio of randomly folded aluminum foils is also universal, *i.e.* independent on the foil thickness, sheet size, and folding contraction ratio. So, our findings suggest that randomly folded elasto-plastic and predominantly plastic sheets belong to different universality classes.

It should be pointed out that the universality of Poisson's index and ratio is very surprising, taking into account that the Poisson's ratio of common porous materials (metals, ceramics, polymers, and soils) strongly depends on the porosity ($p$) or relative mass density ($\rho/\rho_0 = (1-p)$) [15], which in the case of randomly folded sheets depends on the contraction ratio as $\rho/\rho_0 = K^3$ [38].

The findings of this work provide a novel insight into the relation between the topology and mechanical properties of randomly folded matter. We expect that these findings will stimulate the theoretical studies and numerical simulations of the Poisson's effect in the crumpled phase of randomly folded thin sheets and polymerized membranes.

**Acknowledgments**

This work has been supported by CONACyT of the Mexican Government under Project No. 44722.

**Figure captions**

**Figure 1**. Experimental set-ups for tests of: a, b) uniaxial and c) radial compressions of randomly folded paper (a, c) and aluminium foil (b).

**Figure 2**. a) Typical force (F) – compression ($\lambda_\parallel$) curve of randomly folded paper under the uniaxial compression: loading (1), unloading (2), relaxation shown in fig 2 d (3); b) typical stress (MPa) – strain (arbitrary units) behaviour of randomly folded aluminium foil under uniaxial compression: loading (1), unloading (2); c) compression force $-F$ (N) versus time $t$ (hours) for fixed compression ratios $\lambda_\parallel = 0.5$; and d) lateral expansion ratio $\lambda_\perp$ versus time $t$ (hours) for fixed compression ratios $\lambda_\parallel = 0.8$ (1) 0.5 (2), and 0.3 (3).

**Figure 3**. a) Perimeter $P$ (mm) versus height $H = \lambda_\parallel H_0$ for axially compressed cylinders folded from sheets of paper of thickness $h = 0.039$ mm and size $L = 500$ mm with different contraction ratios $K = 3.9$ (1), 5.6 (2), and 8.3 (3). b) Lateral expansion ratio $\lambda_\perp$ versus axial compression ratio $\lambda_\parallel$ for all folded paper sheets tested under uniaxial compression; open symbols correspond to cylinders folded from sheets of size 250 mm and full symbols – to cylinders folded from sheets of size 500 mm from papers of thickness: 0.024 (1,2), 0.030 (3,4), 0.039 (5,6), 0.068 (7,), and 0.087 mm (9,10) with different contraction ratios; solid line – data fitting with the scaling relations (2) with $\nu = 0.17$ (insert shows the same graph plotted in log-log coordinates); d) axial expansion ratio $\lambda_H$ versus radial compression ratio $\lambda_R$ for radially compressed cylinders folded from sheets of paper of size $L = 500$ mm and thickness 0.024 (1, 2) and 0.068 mm (3, 4)

with contraction ratios (1, 3), (2, 4); symbols – experimental data, solid line – data fitting by the scaling relations (4) with $\alpha = 0.2$.

**Figure 4**. a) Lateral expansion ratio $\lambda_\perp$ versus axial compression ratio $\lambda_\parallel$ for randomly folded aluminum foil of size 500x500 mm$^2$; circles – experimental data, doted line – data fitting with equations (3) with $\nu = 0.22$ ($R^2 = 0.981$), solid line – data fitting with equations (1) with $\nu_e = 0.34$ ($R^2 = 0.998$). b) Lateral strains $-\varepsilon_\perp = (P - P(0))/P(0)$ versus axial strains $\varepsilon_\parallel = (R_0 - R)/R_0$ for all aluminum foils tested in this work; solid line – data fitting with $\nu_e = 1/3$ ($R^2 = 0.974$).

[35] In the case of sheets tested immediately after folding (without 7 days of relaxation) we found that the perimeter of compressed cylinder increases about 10% during 6 days.

[36] We were not able to perform the radial compression tests of folded aluminum foils.

[37] Mechanical properties of papers are very sensible to air humidity and temperature.

[38] In the case of randomly folded sheets under hydrostatic compression, the scaling relation $R \propto F^{-\delta}$ with $\delta = 1/4$ [8] implies that the bulk modulus scales with the mass density of folded configuration $\rho \propto M/R^3$ as $B \propto R^3 \partial(F/R^2)/\partial R^3 \propto \rho^n \propto (1-p)^n \propto K^{3n}$ with the universal elastic exponent $n = (1+2\delta)/3\delta = 2$.

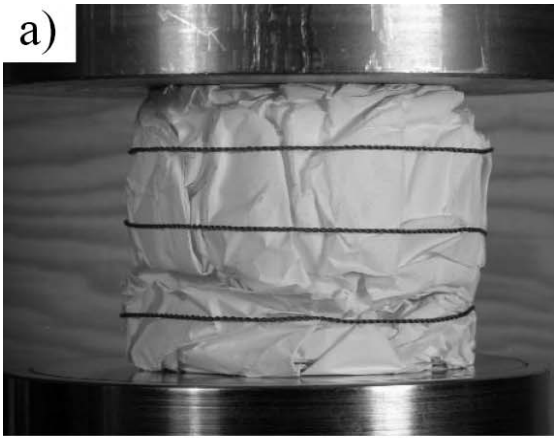
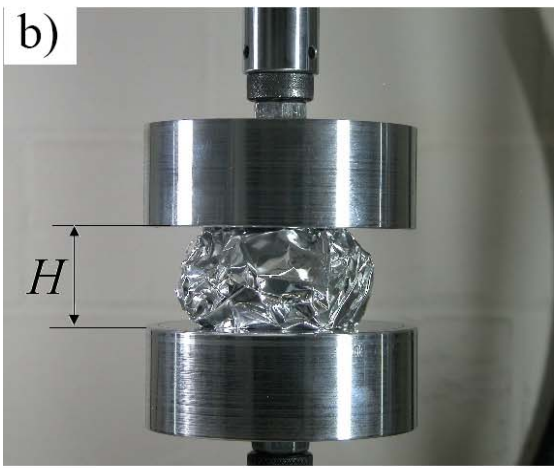
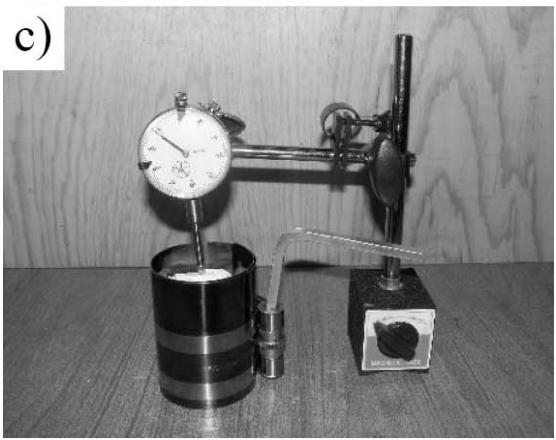

Figure 1

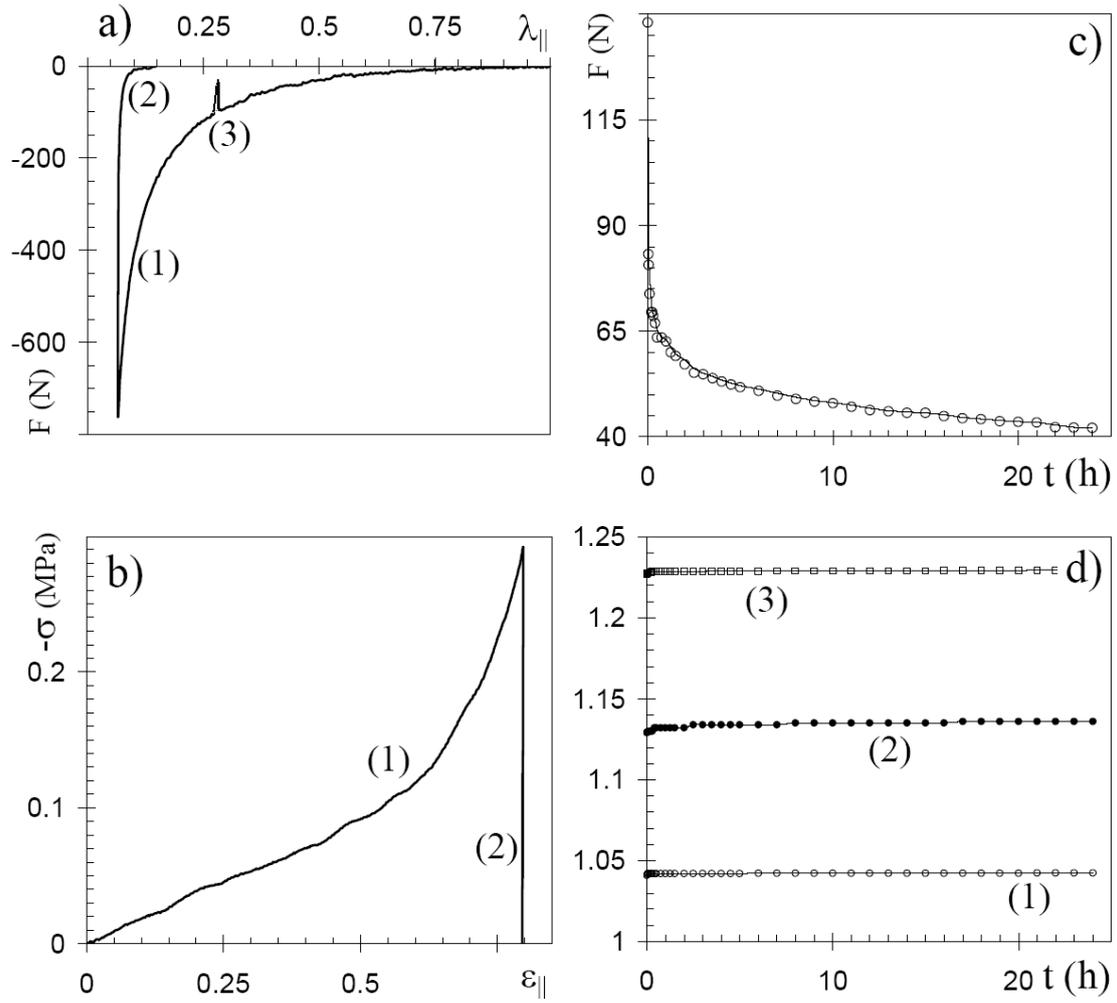

Figure 2

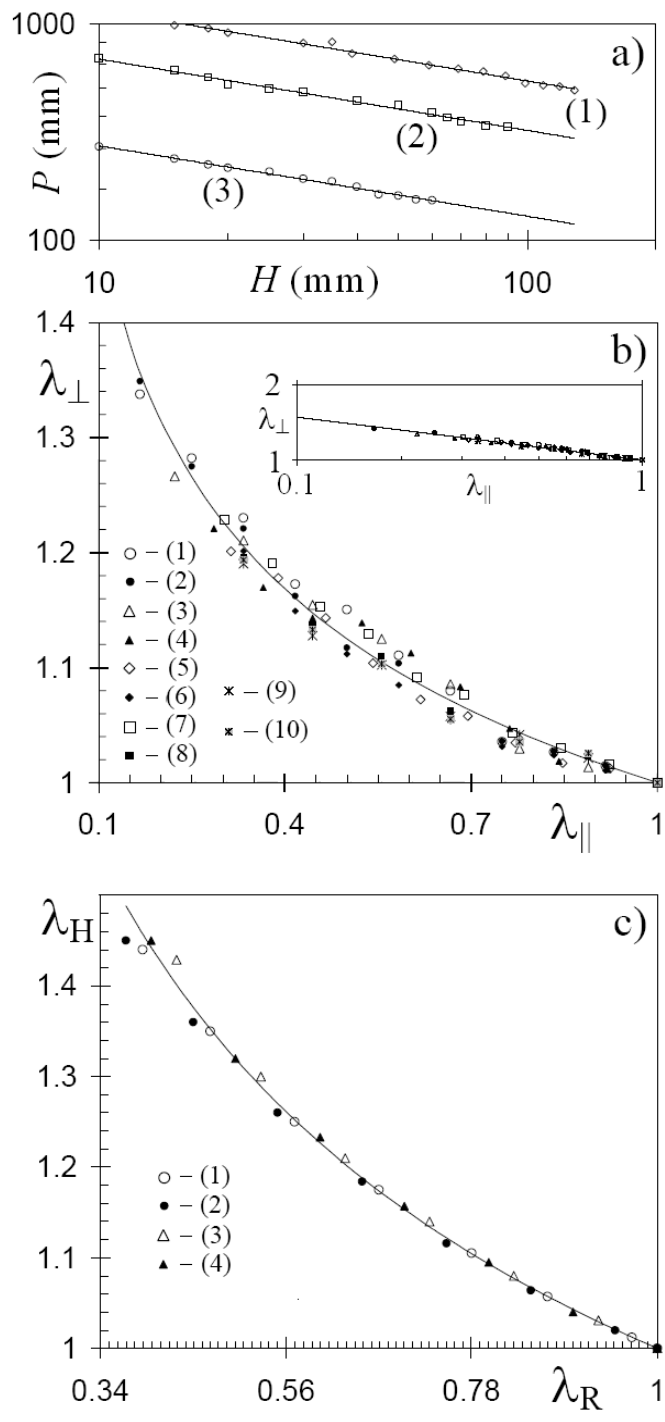

Figure 3.

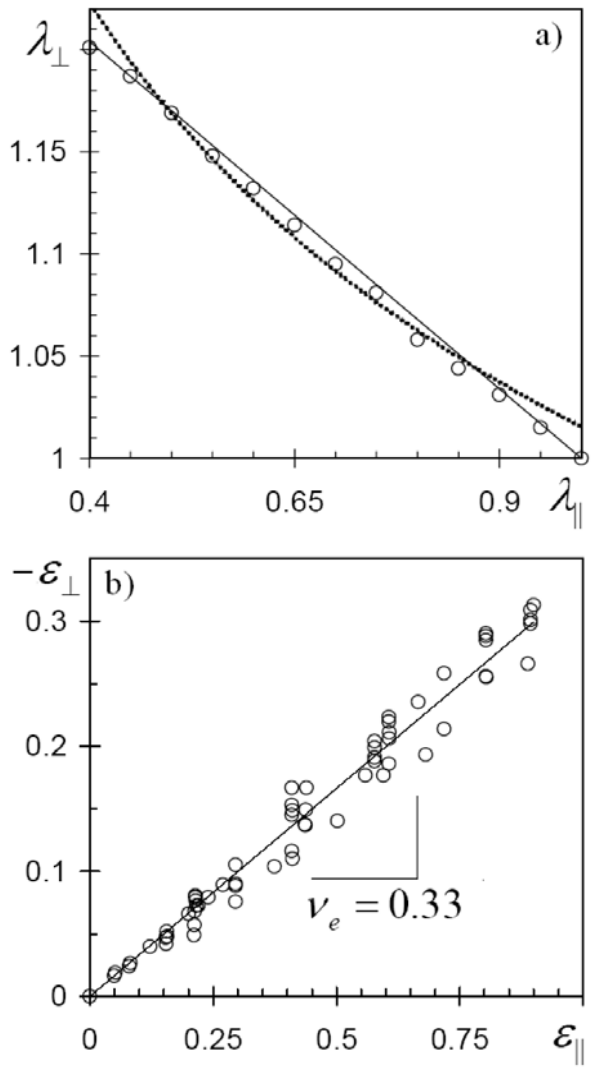

Figure 4.